\documentclass[aps,prb,preprint]{revtex4}
\usepackage{graphicx}

\begin{document}

\title{Superconductivity in Fluorine-Arsenide Sr$_{1-x}$La$_x$FeAsF}

\author{Xiyu Zhu, Fei Han, Peng Cheng, Gang Mu, Bing Shen and Hai-Hu Wen}\email{hhwen@aphy.iphy.ac.cn }

\affiliation{National Laboratory for Superconductivity, Institute of
Physics and Beijing National Laboratory for Condensed Matter
Physics, Chinese Academy of Sciences, P. O. Box 603, Beijing 100190,
China}

\begin{abstract}
Since the discovery of superconductivity\cite{1} at 26 K in
oxy-pnictide $LaFeAsO_{1-x}F_x$, enormous interests have been
stimulated in the fields of condensed matter physics and material
sciences. Among the five different structures in this broad type of
superconductors\cite{2,3,4,5,6}, the ZrCuSiAs structure has received
special attention since the $T_c$ has been quickly promoted to 55-56
K\cite{7,8,9,10,11} in fluorine doped oxy-pnictides REFeAsO (RE =
rare earth elements). The superconductivity can also be induced by
applying a high pressure to the undoped samples\cite{12,13}. The
mechanism of superconductivity in the FeAs-based system remains
unclear yet, but it turns out to be clear that any change to the
structure or the building blocks will lead to a change of the
superconducting transition temperatures. In this Letter, we report
the fabrication of the new family of compounds, namely
fluorine-arsenides DvFeAsF (Dv = divalent metals) with the ZrCuSiAs
structure and with the new building block DvF instead of the REO
(both the layers DvF and REO have the combined cation state of
"+1"). The undoped parent phase has a Spin-Density-Wave like
transition at about 173 K for SrFeAsF, 118 K for CaFeAsF and 153 K
for EuFeAsF. By doping electrons into the system the resistivity
anomaly associated with this SDW transition is suppressed and
superconductivity appears at 32 K in the fluorine-arsenide
Sr$_{1-x}$La$_x$FeAsF (x = 0.4). Our discovery here initiates a new
method to obtain superconductors in the FeAs-based system.
\end{abstract}
\maketitle

By using the solid state reaction method, we successfully fabricated
the fluorine-arsenide material DvFeAsF (Dv = Sr, Eu and Ca) and La
doped Sr$_{1-x}$La$_x$FeAsF (x = 0.0 - 0.4). The detailed
fabrication process for the samples is given in Methods. It is known
that the F-doped REFeAsO superconductors (RE = rare earth elements)
have a ZrCuSiAs structure (see inset in Figure 1) with the
alternative building series of RE$^{3+}$O$^{2-}$ and
Fe$^{2+}$As$^{3-}$ and the RE$^{3+}$O$^{2-}$ has a total cation
state of "+1". It thus becomes quite straightforward that one may
try different building blocks with the same cation state of "+1" to
substitute REO. The combination of Dv$^{2+}$F$^{-}$ with Dv the
divalent elements (Ca, Sr, Eu, etc.) may be a good choice. Actually
the SrF based parent compound was fabricated in the copper-sulfur
system SrFCuS\cite{14,15} and very recently in the FeAs-based
systems\cite{Tegel,HanPRB}. Superconductivity at 22 K was found in
CaFe$_{1-x}$Co$_x$AsF\cite{Hosono1} and 4 K in
SrFe$_{1-x}$Co$_x$AsF\cite{Hosono2}. In the main panel of Figure 1
we present the x-ray diffraction patterns of the compound of SrFeAsF
and the La-doped sample Sr$_{1-x}$La$_x$FeAsF, together with the
parent phases of CaFeAsF and EuFeAsF. Taking the SrFeAsF as the
example, it is clear that almost all main peaks can be indexed by a
tetragonal structure with a = b = 4.004 $\AA$ and c = 8.971 $\AA$
for the parent phase, and a = b = 3.976 $\AA$ and c = 8.960 $\AA$
for the doped samples. Comparing with the LaFeAsO samples, the a- or
b- axes contract a bit but the c-axis lattice constant expands a
lot. This is understandable since the radius of Sr$^{2+}$ is 1.12
$\AA$ which is larger than that of La$^{3+}$ (1.06 $\AA$), although
the radius of F$^{-}$ is 1.33 $\AA$ which is comparable to that of
O$^{2-}$ (1.32 $\AA$). Therefore it is confident that the dominant
component in our sample comes from Sr$_{1-x}$La$_x$FeAsF. There are
some tiny peaks marked by the asterisks which may come from the
impurity phase of SrF$_2$, this can be easily formed during the
synthesizing process and should be avoided. Data in Figure 1 clearly
show that the parent phases with Sr, Ca and Eu based on SrF can be
formed.

In Figure 2 we show the temperature dependence of resistivity of the
undoped samples DvFeAsF (Dv = Sr, Eu and Ca). For the undoped
samples, a clear resistivity anomaly was observed at about 118 K for
CaFeAsF, 173 K for SrFeAsF and 153 K for EuFeAsF. Here we take the
case of SrFeAsF for further discussion. Above 173 K the resistivity
increases slightly with decreasing temperature, but it drops sharply
below 173 K. This resistivity anomaly was found to be corresponding
very well to a magnetic anomaly\cite{HanPRB} measured by DC
magnetization and was attributed to the formation of a
Spin-Density-Wave order or a structural transition in
REFeAsO.\cite{16,17} By applying a magnetic field, a strong
magnetoresistance was observed below 173 K (not shown here), which
appeared also in the oxy-pnictide systems REFeAsO .\cite{ChengP} The
overall behavior of resistivity mimics that of the LaFeAsO, which
again indicates that our samples here are parent phases with the
ZrCuSiAs structure and the FeAs layers as that in oxy-pnictides.

By doping electrons into the parent phase, we observed
superconductivity in Sr$_{1-x}$La$_x$FeAsF. In Fig.3 we present the
temperature dependence of resistivity for three samples with the
nominal composition x=0, 0.2 and 0.4, respectively. The XRD data of
the samples x=0 and 0.2 were shown in Fig.1. As one can see, when
electrons are doped into the system, the SDW anomaly at about 173 K
is suppressed and it moves to about 60 K at a nominal composition of
0.2. Meanwhile the superconductivity at about 25 K (onset) appears.
For the doped sample Sr$_{0.6}$La$_{0.4}$FeAsF a superconducting
transition occurs at about 32 K, as indicated by the arrow where the
resistivity starts to drop away from the normal state background.
The state of zero resistance is achieved at about 24.6 K. It is
interesting to note that the transition temperature here (32 K) is
already much higher than the maximum $T_c$ = 4 K\cite{Hosono2} as
recently observed in SrFe$_{1-x}$Co$_x$AsF. This strongly suggests
that doping to the sites of Sr is more efficient to get
superconductors with higher T$_c$ since the FeAs-planes which is
assumed for superconductivity remain unchanged. Occasionally we even
see superconductivity at temperatures higher than 32 K, but those
samples exhibit quite some impurities, therefore we leave it to a
future report. The resistivity of the sample with x = 0.4  shows a
roughly linear behavior in the normal state, which looks rather
similar to that in the electron doped sample
LaFeAsO$_{0.9}$F$_{0.1}$\cite{ZhuXY} and is in sharp contrast to the
hole doped sample La$_{1-x}$Sr$_x$FeAsO\cite{WenEPL}. This bulk
superconductivity is also proved by the magnetic measurements, as
shown in Figure 4. One can see that a diamagnetic transition occurs
at about 27 K, which corresponds to the middle transition
temperature of the resistivity data. We believe that the
supercoducting transition temperature will be improved higher when
the synthesizing condition is further optimized. It is important to
note that the superconductivity in the present sample is certainly
not originated from the La-doped $SrFe_2As_2$ or the F-doped
LaFeAsO. In the former case, as far as we know, no superconductivity
was found in La-doped SrFe$_2$As$_2$. In addition, no any peaks of
SrFe$_2$As$_2$ can be observed from the XRD data in our present
doped samples. In the latter case, one needs oxygen to form the
LaFeAsO phase which is limited to below 0.1 ppm in our case by using
glove box. Another reason to rule out this possibility is that the
resistivity anomaly (the structural transition and/or SDW) occurs at
173 K in the parent phase SrFeAsF, which is much higher than that in
LaFeAsO (around 150 K), but much lower than that in SrFe$_2$As$_2$
(around 205 K).\cite{SrK}

In summary, by using the combination of DvF, we fabricated the new
family of fluorine-arsenide materials DvFeAsF (Dv = Sr, Eu and Ca)
with the ZrCuSiAs structure. In all three cases, the parent phase
exhibits a resistivity anomaly (118 K for Ca, 173 K for Sr and 153 K
for Eu) which is related to the structural/SDW transition. By
partially substituting Sr with La, we observed superconductivity in
Sr$_{1-x}$La$_x$FeAsF (x=0.2, 0.4). As a typical example here we
show the superconductivity at 32 K in Sr$_{0.6}$La$_{0.4}$FeAsF.
Using different combinations of the divalent elements (Sr, Eu, Ca,
etc.) with different dopants in this fluorine-arsenide family, new
superconductors are expected to be produced.

\section{Methods}
\subsection{Sample preparation.}
The polycrystalline samples were synthesized by using a two-step
solid state reaction method.\cite{ZhuXY} First LaAs and SrAs (or
CaAs, EuAs etc.) powders were obtained by the chemical reaction
method with La grains (purity 99.99\%), Sr (or Ca, Eu etc.) pieces
and As grains. Then they were mixed with FeF$_3$ (purity 99\%) and
Fe powder (purity 99.99\%) in the formula Dv$_{1-x}$La$_x$FeAsF
(Dv=Sr, Eu and Ca), ground and pressed into a pellet shape. All the
weighing, mixing and pressing procedures were performed in a glove
box with a protective argon atmosphere (both H$_2$O and O$_2$ are
below 0.1 PPM). The pellet was sealed in a quartz tube with 0.2 bar
of Ar gas and followed by heat treatment at 950 $^o$C for 60 hours.
Then it was cooled down slowly to room temperature. Sometimes the
resultant pellet was ground again, sealed in a quartz tube for a
second sintering at 1000 $^o$C.

\subsection{Measurements.}
The DC magnetic measurements were done with a superconducting
quantum interference device (Quantum Design, SQUID, MPMS7). The
zero-field-cooled magnetization was measured by cooling the sample
at zero field to 2 K, and the data were collected during the warming
up process. The resistivity measurements were done with a physical
property measurement system (Quantum Design, PPMS9T) with a
six-probe technique. The current direction was changed for measuring
each point in order to remove the contacting thermal power.

This work was supported by the Natural Science Foundation of China,
the Ministry of Science and Technology of China (973 Projects
No.2006CB601000, No. 2006CB921802), and Chinese Academy of Sciences
(Project ITSNEM). We thank Lei Fang for some helps in making Figure
1.

\section{Competing financial interests}

The authors declare that they have no competing financial interests.

\section{Author Contributions}

HHW designed and coordinated the major part of the experiment,
analyzed the data and wrote the paper. XYZ and FH made samples of
Sr$_{1-x}$La$_x$FeAsF and EuFeAsF and did the experiment on them; PC
and BS did for CaFeAsF; GM helped in analyzing data and correcting
the paper.

\section{Author information}

Correspondence and requests for materials should be addressed to HHW
(hhwen@aphy.iphy.ac.cn)

\newpage
\begin{figure}
\includegraphics[width=14cm]{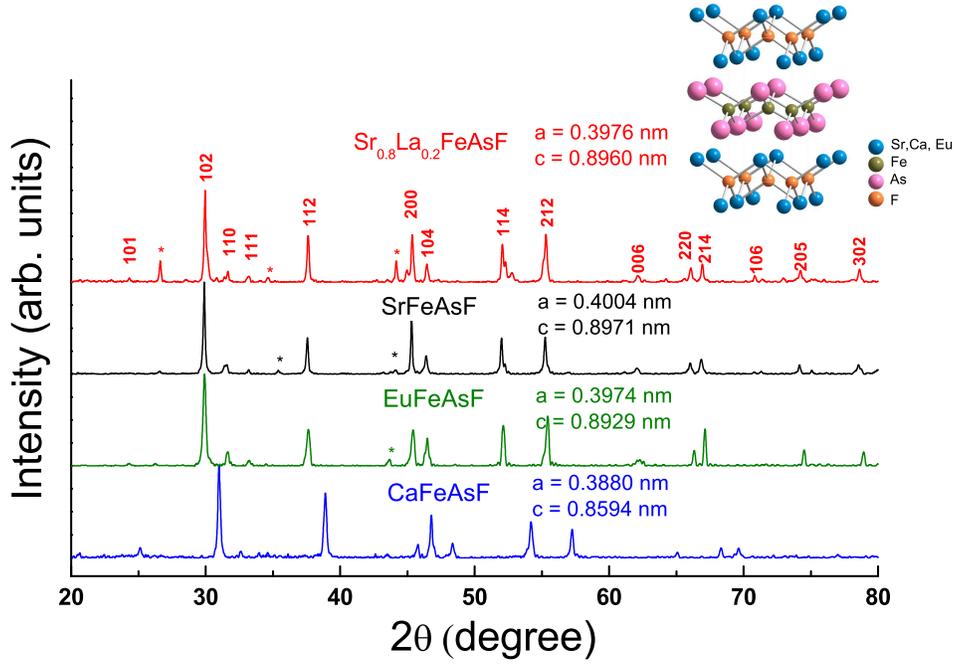}
\caption{(Color online) X-ray diffraction patterns of the samples
Sr$_{1-x}$La$_x$FeAsF (x=0.0 and 0.20). All main peaks can be
indexed by a tetragonal structure with a = b = 4.004 $\AA$ and c =
8.971 $\AA$ for the undoped sample SrFeAsF, a = b = 3.976 $\AA$ and
c = 8.960 $\AA$ for the doped sample Sr$_{0.8}$La$_{0.2}$FeAsF. This
indicates that the dominant phase here is Sr$_{1-x}$La$_x$FeAsF.
Shown together are the X-ray diffraction patterns of the parent
phases EuFeAsF and CaFeAsF. The asterisks mark the peaks from the
tiny impurity phase SrF$_2$.} \label{fig1}
\end{figure}

\begin{figure}
\includegraphics[width=14cm]{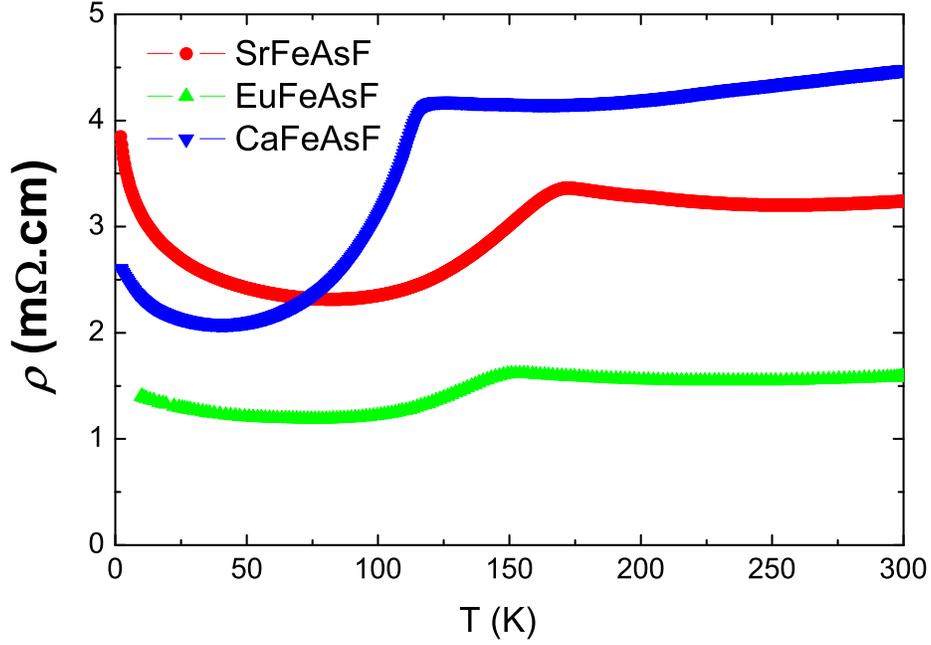}
\caption{(Color online) Resistivity versus temperature in SrFeAsF,
CaFeAsF and EuFeAsF. For the three undoped (parent) samples, the
resistivity anomaly occurs at different temperatures: 118 K for Ca,
173 K for Sr and 153 K for Eu. This anomaly may be attributed to the
structural transition from tetragonal (high temperature) to
orthorhombic (low temperature) phase or the SDW transition. The
exact reason for the correlation between the anomaly temperature and
the radii of the divalent metals remains unclear. } \label{fig2}
\end{figure}

\begin{figure}
\includegraphics[width=14cm]{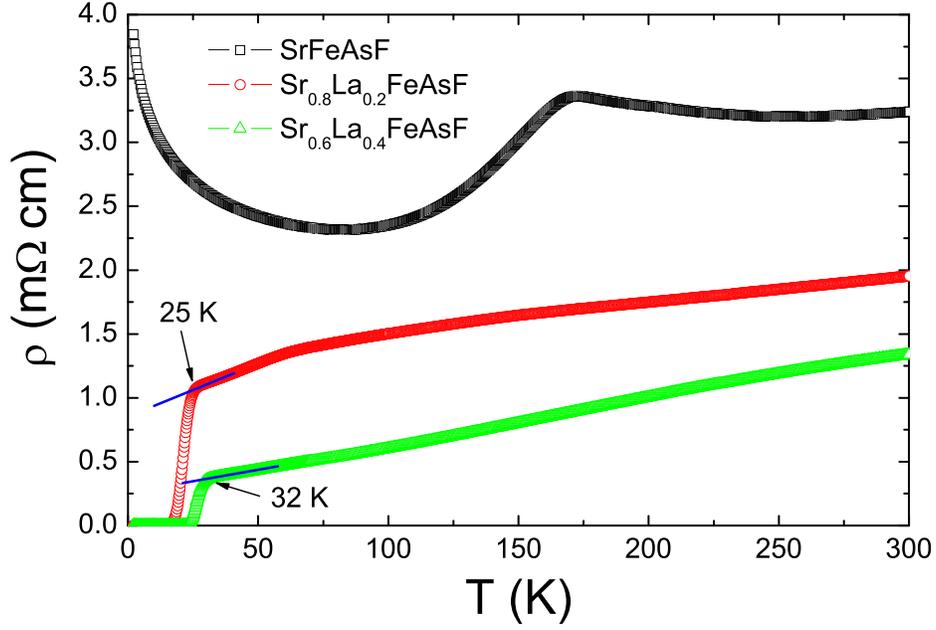}
\caption{(Color online)  The temperature dependence of resistivity
for three samples Sr$_{1-x}$La$_{x}$FeAsF (x=0, 0.2 and 0.4). One
can see that the temperature related to the resistivity anomaly is
suppressed to about 60 K on the sample with nominal composition of x
= 0.2. Meanwhile the superconductivity appears at about 25 K
(onset). For the sample with x=0.4, a clear superconducting
transition occurs at about 32 K (onset), and the state with zero
resistivity is achieved at about 24.6 K.  } \label{fig3}
\end{figure}

\begin{figure}
\includegraphics[width=14cm]{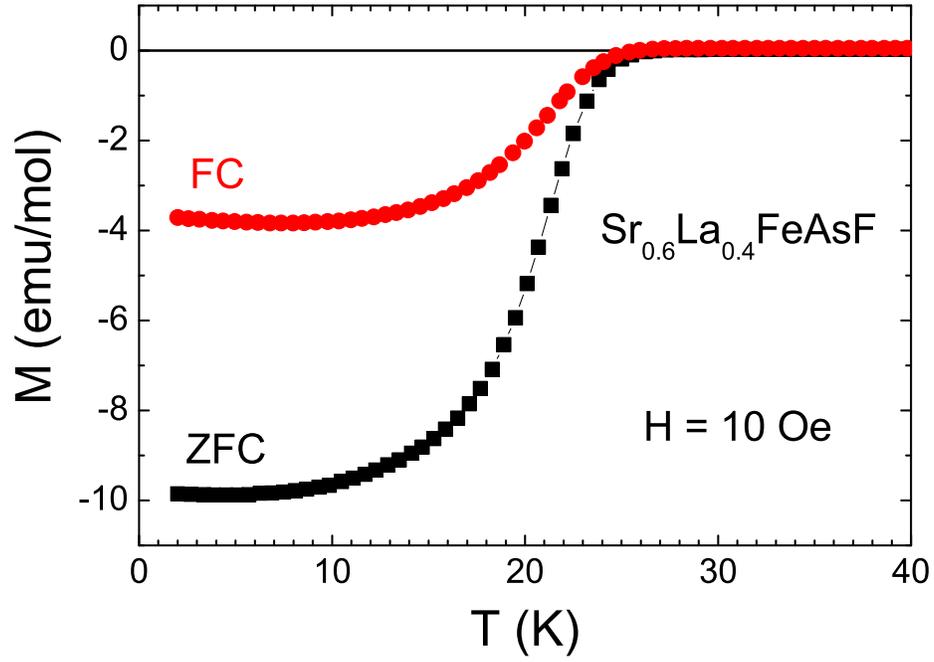}
\caption{(Color online)  The temperature dependence of DC
magnetization of the sample Sr$_{0.6}$La$_{0.4}$FeAsF measured at a
DC field of 10 Oe. This clearly indicates a bulk superconductivity
in the sample Sr$_{0.6}$La$_{0.4}$FeAsF. } \label{fig4}
\end{figure}
\end{document}